# An Alternative Method for Solving Security-Constrained Unit Commitment with Neural Network Based Battery Degradation Model

Cunzhi Zhao
*Student Member, IEEE*
Department of Electrical and Computer Engineering
University of Houston
Houston, TX, USA
czhao20@uh.edu

Xingpeng Li
*Senior Member, IEEE*
Department of Electrical and Computer Engineering
University of Houston
Houston, TX, USA
xli82@uh.edu

*Abstract*-- **Battery energy storage system (BESS) can effectively mitigate the uncertainty of variable renewable generation and provide flexible ancillary services. However, degradation is a key concern for rechargeable batteries such as the most widely used Lithium-ion battery. A neural network based battery degradation (NNBD) model can accurately quantify the battery degradation. When incorporating the NNBD model into security-constrained unit commitment (SCUC), we can establish a battery degradation based SCUC (BD-SCUC) model that can consider the equivalent battery degradation cost precisely. However, the BD-SCUC may not be solved directly due to high non-linearity of the NNBD model. To address this issue, the NNBD model is linearized by converting the nonlinear activation function at each neuron into linear constraints, which enables BD-SCUC to become a linearized BD-SCUC (L-BD-SCUC) model. Case studies demonstrate the proposed L-BD-SCUC model can be efficiently solved for multiple BESS buses power system day-ahead scheduling problems with the lowest total cost including the equivalent degradation cost and normal operation cost.**

*Index Terms*— **Battery degradation, Battery energy storage system, Bulk power system, Energy management system, Machine learning, Security constrained unit commitment, Day-ahead scheduling, Neural network, Optimization.**

## Nomenclature

### Sets

| | |
|---|---|
| $G$ | Set of generators. |
| $S$ | Set of battery energy storage systems. |
| $T$ | Set of time intervals. |
| $K$ | Set of lines. |
| $K(n+)$ | Set of lines with bus $n$ as receiving bus. |
| $K(n-)$ | Set of lines with bus $n$ as sending bus. |
| $N$ | Set of buses. |

### Indices

| | |
|---|---|
| $g$ | Generator g. |
| $k$ | Line k. |
| $t$ | Time interval t. |
| $n$ | Bus n. |
| $s$ | Battery energy storage system s. |

### Parameters

| | |
|---|---|
| $c_g$ | Linear cost of generator g. |
| $c_g^{NL}$ | No load cost for generator g. |
| $c_g^{SU}$ | Start-up cost for generator g. |
| $\Delta T$ | Length of a single dispatch interval. |
| $P_g^{max}$ | Maximum output power of generator g. |
| $P_g^{min}$ | Minimum output power of generator g. |
| $P_g^{Ramp}$ | Ramping limit of generator g. |
| $P_s^{max}$ | Maximum charge/discharge power of BESS s. |
| $P_s^{min}$ | Minimum charge/discharge power of BESS s. |
| $P_k^{max}$ | Long-term thermal line limit for line k. |
| $P_d^t$ | Load demand at time period t. |
| $b_k$ | Susceptance of line k. |
| $E_s^{Initial}$ | Initial energy capacity of BESS s. |
| $\eta_s^{Disc}$ | Discharge efficiency of BESS s. |
| $\eta_s^{Char}$ | Charge efficiency of BESS s. |

### Variables

| | |
|---|---|
| $P_g^t$ | Output of generator g in time period t. |
| $U_g^t$ | Commitment status of generator in time period t. |
| $V_g^t$ | Start-up variable of generator in time period t. |
| $P_k^t$ | Flow on line $k$ in time period t. |
| $P_t^t$ | Available renewable power in time period t. |
| $P_{Disc}^{t,s}$ | Discharging power of BESS s in time period t. |
| $P_{Char}^{t,s}$ | Charging power of BESS s in time period t. |
| $U_{Disc}^{t,s}$ | Discharging status of BESS s determined at time period t. It is 1 if discharging status; otherwise 0. |
| $U_{Char}^{t,s}$ | Charging status of BESS s determined at time period t. It is 1 if charging status; otherwise 0. |
| $E_s^t$ | Energy storage capacity at time period t. |
| $\theta_n^t$ | Phase angle of bus n in time period t. |
| $\theta_m^t$ | Phase angle of bus n in time period t. |
| $SOC_s^t$ | State of charge for BESS s in time period t. |

## I. INTRODUCTION

Power systems typically have much higher generation capacity than the peak load to ensure the resource adequacy and grid reliability. The electric grid is inefficient since a large amount of produced electricity is wasted [1]. It may get worse with the worldwide policy of decarbonation by implementing more renewable energy sources (RESs). The stochastic and intermittent generation of high penetration RESs may substantially weaken the system's stability and further reduce the grid efficiency. Fortunately, the uncertainty and inefficiency issues can be addressed by taking advantage of battery energy storage system (BESS) [2]-[3].

BESS consists battery packs that are connected in parallel or series. Lithium-ion battery is widely used in BESS and electric vehicles due to its nature of high energy density and low memory effect. However, Lithium-ion battery will degrade during cycling and it is quite hard to predict the amount of battery degradation [4]-[5]. The lifetime of BESSs resulted by degradation is highly sensitive to dispatch strategies that are related to several critical battery degradation factors. Thus, failing to consider battery degradation accurately may reduce the lifetime and result into financial losses of investors [6].

Plenty of previous studies have developed various battery degradation models for BESS. A piecewise linear battery degradation model that is based on Arrhenius law is proposed in [7] to predict battery degradation. However, the proposed degradation model only reflects the impact of depth of discharge (DOD), which is not sufficient. Similarly, the battery degradation in [8] is calculated based on the remaining useful life (RUL) of BESS that is predicted by the DOD of each cycle. It is not reasonable to consider the linear relationship between DOD and RUL throughout the lifetime of BESS. Moreover, the RUL is affected by several other degradation factors besides the DOD. A linear degradation rate is applied to quantify the battery degradation cost in the optimization problem [9]-[11]. The linear degradation may decrease the difficulty of solving the unit commitment problem, but the inaccurate degradation information of the linear model may substantially reduce the lifetime of the BESS. For all the aforementioned battery degradation models, they either consider a linear degradation cost or the models missing several critical degradation factors such as state of charge (SOC), C rate, and ambient temperature; none of them developed a comprehensive model to cover the majority critical degradation factors.

A data driven degradation model is presented in [12] to predict the degradation. However, DOD and SOC are the only variables in the training dataset, which indicates the model is lack of other critical degradation factors. A neural network based battery degradation (NNBD) is developed in [13]. Although the NNBD model can accurately predict the battery degradation with major degradation factors (SOC, DOD, C rate, state of health (SOH) and ambient temperature), the proposed iterative method seems to only address systems with only one BESS integrated bus and cannot scale to large-scale systems.

Security-constrained unit commitment (SCUC) is one of the most important optimization problems in power system day-ahead scheduling [14]-[18]. Nonetheless, the battery degradation is not considered in those SCUC models. With the increasing installed BESS capacity in the power system, it is very important to give thought to the battery degradation in future SCUC models to make the best use of the BESSs.

To bridge the aforementioned gaps, this paper proposes a novel security-constrained unit commitment model with linearized neural network based battery degradation model, referred to as linearized battery degradation model based SCUC (Linearized BD-SCUC or L-BD-SCUC). The proposed L-BD-SCUC model that considers the equivalent battery degradation cost is directly solvable. The NNBD model is structured to learn and predict the value of battery degradation with major degradation factors. The non-linear activation function for each neuron in each hidden layer is linearized to enable a linearized NNBD model. As a result, L-BD-SCUC can be solved directly to provide the optimal solution with the lowest total cost that is the sum of the operation cost and the equivalent battery degradation cost.

The remainder of the paper is organized as follows. The mathematical formulation for the traditional SCUC model is presented in Section II. Section III explains the proposed L-BD-SCUC model. Case studies and discussions are presented in Section IV. Section V concludes the paper.

## II. TRADITIONAL SCUC MODEL

A traditional SCUC (T-SCUC) model is established as a benchmark model to gauge the proposed L-BD-SCUC model. This T-SCUC model consists of (1)-(18) as described below and it does not consider equivalent degradation cost of BESS. The cost of BESS degradation will be presented in next section. The objective of the T-SCUC model is to minimize the total operation cost of the generators which is shown below:

$$f(cost) = f^G \tag{1}$$

where $f^G$ denotes the total cost of all the generator units as defined in (2).

$$f^G = \sum_{t \in S_T} \sum_{g \in S_G} P_g^t c_g + U_g^t c_g^{NL} + V_g^t c_g^{SU} \tag{2}$$

The nodal power balance equation involving synchronous generators, renewable energy sources, BESSs and demand of bus $n$ is shown in (3). Constraints (4-6) represent the power output limits and ramping limits of each generators. Equations (7)-(9) define the relation between generator start-up status and generator on/off status. The thermal limit of the transmission line is enforced by (10). Constraint (11) represents the network power flow equation. As shown in (12), the SOC level can be represented by the ratio between the current stored energy and maximum available energy capacity. Constraints (13)-(14) enforce the charging/discharging power limits of BESS. Constraint (15) restricts the BESS to be either in charging mode or in discharging mode or stay idle. Equation (16) calculates the stored energy of BESS at each time interval. The ending BESS SOC level is forced to equal the initial value in (17). Equation (18) enforces the limit of the stored energy for BESS.

$$\sum_{g \in S_G} P_g^t + \sum_{r \in S_R} P_r^t + \sum_{k \in K(n-)} P_k^t + \sum_{s \in S_S} P_{Disc}^{t,S} = \sum_{k \in K(n+)} P_k^t + \sum_{d \in S_L} P_d^t + \sum_{s \in S_S} P_{Char}^{t,S}, \forall n, t, \tag{3}$$

$$P_g^{Min} U_g^t \le P_g^t \le P_g^{Max} U_g^t, \forall g, t, \tag{4}$$

$$P_g^{t+1} - P_g^t \le \Delta T \cdot P_g^{Ramp}, \forall g, t, \tag{5}$$

$$P_g^t - P_g^{t+1} \le \Delta T \cdot P_g^{Ramp}, \forall g, t, \tag{6}$$

$$V_g^t \ge U_g^t - U_g^{t-1}, \forall g, t, \tag{7}$$

$$V_g^{t+1} \le 1 - U_g^t, \forall g, t, \tag{8}$$

$$V_g^t \le U_g^t, \forall g, t, \tag{9}$$

$$-P_k^{Max} \le P_k^t \le P_k^{Max}, \forall k, t, \tag{10}$$

$$P_k^t - b_k(\theta_n^t - \theta_m^t) = 0, \forall k, t, \tag{11}$$

$$SOC_s^t = E_s^t / E_s^{Max}, \forall s, t, \tag{12}$$

$$U_{Char}^{t,s} P_s^{Min} \leq P_{Char}^{t,s} \leq U_{Char}^{t,s} P_s^{Max}, \forall s, t, \tag{13}$$

$$U_{Disc}^{t,s} P_s^{Min} \leq P_{Disc}^{t,s} \leq U_{Disc}^{t,s} P_s^{Max}, \forall s, t, \tag{14}$$

$$U_{Disc}^{t,s} + U_{s}^{t,s} \leq 1, \forall s, t, \tag{15}$$

$$E_s^t - E_s^{t-1} + \Delta T \cdot \left( P_{Disc}^{t-1,s} / \eta_s^{Disc} - P_{Char}^{t-1,s} \eta_s^{Char} \right) = 0, \forall s, t, \tag{16}$$

$$E_s^{t=24} = E_s^{Initial} \ (i \in S_S), \forall s, t, \tag{17}$$

$$E_s^{Min} \leq E_s^t \leq E_s^{Max}, \forall s, t, \tag{18}$$

## III. Linearization of Learning Based Battery Degradation Model and its Incorporation in SCUC

### A. NNBD Model

A fully connected neural network is constructed to model the battery degradation as shown in Fig. 1 [13]. There are two hidden layers and each neuron in those two layers employs the "relu" activation function in the NNBD model. Five critical degradation factors (ambient temperature, c rate, state of charge, depth of discharge and state of health) form a five-element input vector for the neural network. Each input vector corresponds with a single output value which is the amount of battery degradation in percentage respect to the SOH level for the same operation cycle. The training data is generated by the battery aging tests that are simulated by the MATLAB Simulink under different degradation factors. The NNBD model can learn and predict the battery degradation accurately, which is adopted to calculate the equivalent degradation cost in (15), where $c_{BESS}^{Capital}$ and $c_{BESS}^{SV}$ represent the capital investment cost and salvage value of BESS respectively; $SOH_{EOL}$ denotes the state of health value that is considered as end of battery life.

$$f^{BESS} = \frac{c_{BESS}^{Capital} - c_{BESS}^{SV}}{1 - SOH_{EOL}} \sum_t NN(Variables) \tag{15}$$

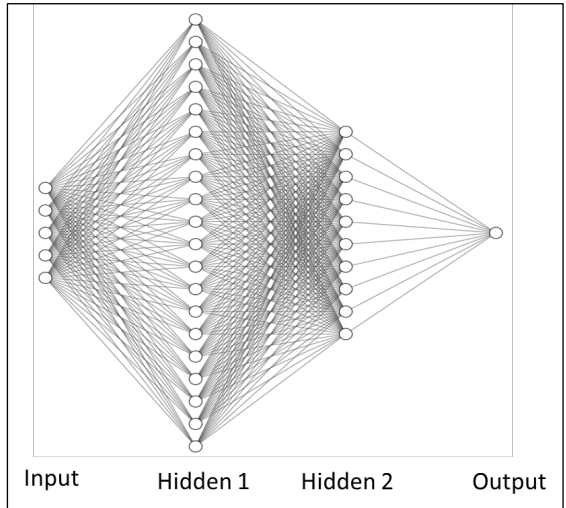

Figure 1. Structure of the NNBD model [13].

### B. Linearization of NNBD model

Since the iteration method developed in [13] for solving BD-SCUC is limited to single-BESS systems, an alternative method L-BD-SCUC is proposed in this paper to linearize the NNBD model and make the BD-SCUC directly solvable.

The proposed L-BD-SCUC model consists of (2)-(17) and (19)-(22). Its objective function is defined in (16). Besides the system operation cost $f^G$, the equivalent battery degradation cost $f^{BESS}$ as shown in (15) is included.

$$f(cost) = f^G + f^{BESS} \tag{16}$$

The NNBD model can be expressed by a set of equations that represent neuron's calculation and activation. Equation (17) represents the calculation for each neuron that involves the input features from the first layer, corresponding weights matrix $W$ and the biases matrix. The non-linear "relu" activation function is represented in (18) which is linearized with an auxiliary variable $\delta_h^i$ by (19)-(22). Note that $\delta_h^i$ is a binary variable: one indicates activation is enabled and zero otherwise. $a_h^i$ represents the activated value of $x_h^i$.

$$x_h^i = \sum x_{h-1}^i * W + Bias \tag{17}$$

$$a_h^i = relu(x_h^i) = max(0, x_h^i) \tag{18}$$

$$a_h^i \leq x_h^i + BigM * (1 - \delta_h^i) \tag{19}$$

$$a_h^i \geq x_h^i \tag{20}$$

$$a_h^i \leq BigM * \delta_h^i \tag{21}$$

$$a_h^i \geq 0 \tag{22}$$

## IV. Case Studies

A typical IEEE 24-bus system [14] that has 33 generators is used as a test bed to evaluate the proposed L-BD-SCUC method in this paper. Fig. 2 illustrates the IEEE 24-bus system. The benchmark model T-SCUC does not consider the battery degradation. The L-BD-SCUC model and the T-SCUC model are solved by the python package "Pyomo" [19] and "Gurobi" optimizer solver [20].

A verification test is first conducted by solving SCUC for a single BESS integrated system to demonstrate the proposed L-BD-SCUC model against the BD-SCUC model. The BESS schedule profile that obtained from the result of L-BD-SCUC model is fed into the NNBD model to calculate the battery degradation and its equivalent cost. From the results, the battery degradation cost obtained from the trained NNBD model is $14,289.50, while it is $14,289.49 reported directly from the L-BD-SCUC model. The degradation cost between the two models are negligible, which verifies the effectiveness of the proposed linearization model. This is actually expected since the piecewise linearization with an auxiliary binary variable is an exact reformulation of activation function "relu".

Table I presents the parameters of BESSs that are installed at different buses in the IEEE 24-bus system. The energy capacities of different BESSs are different. The BESS numbered four has the largest energy capacity and the highest output power among all five BESSs. Table II presents the wind farms that are integrated in the IEEE 24-bus system. There are five wind farms and each contains different numbers of wind turbines. The capacity for each wind turbine is 200 kW. The wind profile data originally from Pecan Street Dataport [21] are scaled for this study.

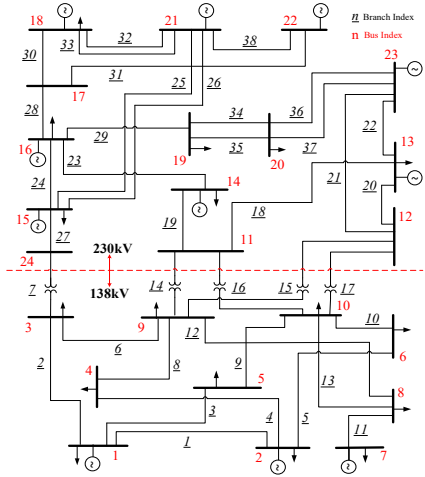

Figure 2. IEEE 24-bus system [14].

Table I BESS parameters.

| BESS No. | Bus No. | Capacity (MWh) | $P\_Max$ (MW) | $P\_Min$ (MW) | Price $/MWh | Initial SOC |
|---|---|---|---|---|---|---|
| 1 | 21 | 50 | 20 | 0 | 100,000 | 40% |
| 2 | 22 | 10 | 4 | 0 | 120,000 | 40% |
| 3 | 7 | 10 | 4 | 0 | 120,000 | 40% |
| 4 | 14 | 200 | 100 | 0 | 75,000 | 40% |
| 5 | 9 | 30 | 10 | 0 | 110,000 | 50% |

Note that $P\_Max$ denotes maximum limit of both charge and discharge power.

Table II Wind farm locations and sizes.

| Wind Farm No. | Wind Farm Bus | # of Wind Turbines |
|---|---|---|
| 1 | 21 | 200 |
| 2 | 22 | 80 |
| 3 | 2 | 100 |
| 4 | 14 | 100 |
| 5 | 15 | 100 |

Table III compares the results from T-SCUC and L-BD-SCUC. The total cost represents the summation of generators' fuel cost and the equivalent battery degradation cost. Since the T-SCUC model does not consider the battery degradation, the equivalent battery degradation cost for T-SCUC in Table II is obtained by collecting the BESS output results from T-SCUC and feed it into the NNBD model. In other words, the battery degradation cost is calculated independently after the T-SCUC is solved while it is directly considered and solved in the L-BD-SCUC model. The results meet the expectation that there will be a total cost reduction achieved by the proposed L-BD-SCUC model. With the proposed L-BD-SCUC model, the total cost decreases by 4.21% comparing with the T-SCUC model. The battery degradation cost significantly decreases by 41.3%. On the other hand, the fuel cost increases by 0.8% due to the change of BESSs' schedule. Comparing with the decreasing battery degradation cost, the increment of fuel cost is insignificant.

Table III Results for IEEE-24 bus system.

| IEEE 24-bus test systems with 5 BESSs | | | |
|---|---|---|---|
| | Fuel Cost ($) | BD Cost ($) | Total Cost ($) |
| T-SCUC | 256,404.60 | 34,643.80 | 291,048.40 |
| L-BD-SCUC | 258,448.90 | 20,348.10 | 278,797.00 |
| Reduction | -0.80% | 41.30% | 4.21% |

Fig. 3 to Fig. 7 present the scheduled BESS operation curves with the L-BD-SCUC model and benchmark model for the five-BESS integrated system. From the results, we can observe that all the BESSs are scheduled to be more "active" for T-SCUC than L-BD-SCUC. "Active" represents the periods that BESS is on charging/discharging status instead of idle. In addition, the BESS output/input power in each time period is generally scheduled to be higher in T-SCUC than L-BD-SCUC. These results are because battery degradation cost is considered in the L-BD-SCUC model. The charge/discharge rate and DOD play a vital role in the NNBD model. Therefore, the BESS is scheduled to charge/discharge in a narrow power range and in less time periods to decrease the amount of battery degradation and the equivalent cost. The results show that the majority time periods are set to idle for BESS 1 and 2 in L-BD-SCUC model. For BESS 3, 4, and 5, active time periods are similar between L-BD-SCUC and T- SCUC. However, they are scheduled in a narrow power range for L-BD-SCUC model to decrease battery degradation. The results may be affected by the bus location of the BESSs and wind farms. Overall, the BESSs' schedule indicates that the proposed L-BD-SCUC method is able to obtain the solution for multi-BESSs integrated bulk power system.

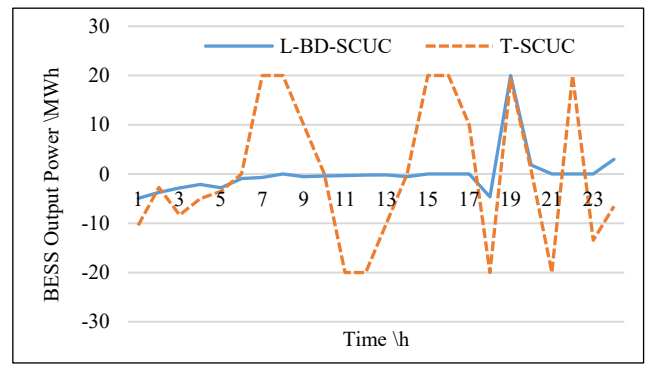

Figure 3. Output power of BESS #1.

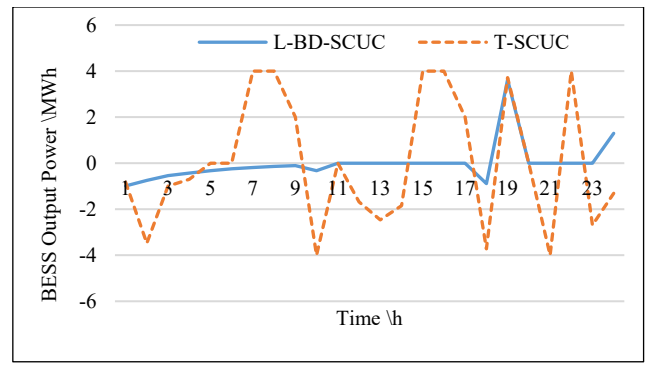

Figure 4. Output power of BESS #2.

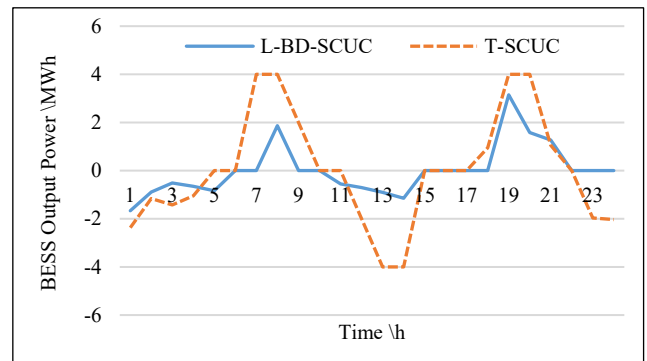

Figure 5. Output power of BESS #3.

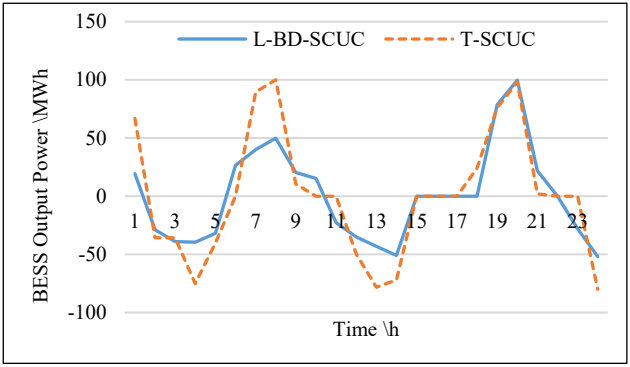

Figure 6. Output power of BESS #4.

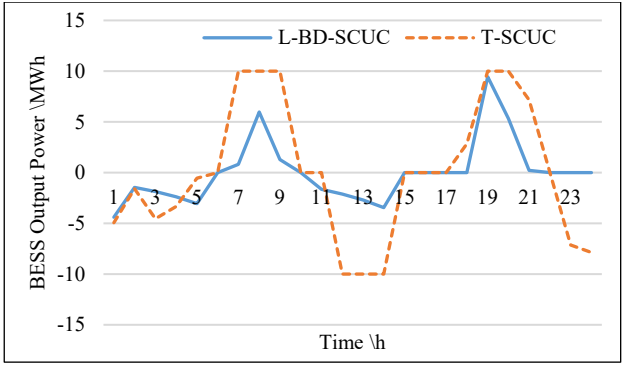

Figure 7. Output power of BESS #5.

Table IV presents the sensitivity tests on optimization relative mipgap (rel-mipgap) with the implement of L-BD-SCUC method shown above. In previous results, we select the 0.01 as the value of rel-mipgap and it took 357.2 seconds to converge. For rel-mipgaps with 0.001 and 0.0001, the program terminates due to the pre-set maximum solving time, which indicates the model cannot find the optimal solution within desired mipgaps. The long computing time and convergence issue motivate us to conduct further sensitivity tests on system scenarios with different numbers of BESSs.

Table IV Relative mipgap tests on 5-BESS system.

| Optimization Mipgap | Total Cost ($) | Degradation Cost ($) | Solving Time (s) |
|---|---|---|---|
| 0.1 | 302,843.2 | 19,515.2 | 47.2 |
| 0.01 | 278,797.0 | 20,348.1 | 357.2 |
| 0.001 | 278,777.4 | 20,338.3 | 3600 |
| 0.0001 | 278,774.5 | 20,338.5 | 3600 |
| 0 | 278,774.5 | 20,338.5 | 3600 |

Table V presents the results of sensitivity analysis on the number of BESSs. The rel-mipgap is set to 0.01 for all the tests in Table V. From Table V, it is clear that the solving time significantly increases as the number of BESSs increases, which indicates the solving efficiency decreases with more BESS buses integrated into the system.

The proposed model can also be applied to illustrate the economic benefits of integrating BESS into power systems. Table VI shows the economic results for different total energy capacities of BESSs. The cost represents the capital investment cost of all the BESSs. The economic benefit represents the total lifetime revenue with the implementation of BESSs. The revenue is calculated based on the difference between the total cost of a BESS integrated system and a benchmark system with no BESS. The expected lifetime is obtained based on average daily battery degradation with $SOH_{EOL}$ being set to 50%. The average daily battery degradation is obtained from the proposed model which consists no extra operation limits for BESS such as cycle limit, rate limit, DOD limit and SOC limit. However, the battery operation is usually limited in order to extend the lifespan in practice. Thus, the actual lifetime should be longer than the results listed in Table VI at the cost of limiting battery daily usage. We find that with a higher installed BESS capacity, the SCUC cost decreases. The SCUC cost here represents the total cost of L-BD-SCUC model. However, the higher BESS capacity may not be the optimal choice. It also depends on the renewable generation capacity and the bus location of BESSs in the system.

Table V Results of sensitivity analysis on number of BESS in the system.

| Numbers of BESS | 1 | 2 | 3 | 4 | 5 |
|---|---|---|---|---|---|
| Solving Time (s) | 17.7 | 29.3 | 134.3 | 234.7 | 357.2 |

Table VI Economic results.

| Total BESS Capacity (MWh) | Cost ($ in millions) | Economic benefit ($ in millions) | Expected lifetime (years) | SCUC total cost ($) |
|---|---|---|---|---|
| 50 | 5.5 | 12.02 | 11.6 | $300,202 |
| 100 | 8 | 46.04 | 11.6 | $292,166 |
| 200 | 15 | 78.65 | 11.4 | $284,136 |
| 300 | 22 | 101.3 | 9.2 | $270,935 |
| 500 | 36 | 102.94 | 9.6 | $273,661 |

## V. CONCLUSION

A novel security-constrained unit commitment model with linearized neural network based battery degradation model is proposed to linearize the learning based battery degradation model to make battery degradation considered SCUC problem directly solvable in this paper. A linearization model is formulated to linearize the activation functions of the NNBD model. The results of this research demonstrate that the proposed L-BD-SCUC method can effectively solve battery degradation-based SCUC for a power system with multiple BESS buses. The statistical economic results of the proposed method give an overview of the potential economic benefit of the BESS integration and provide insights into power system planning. One finding from this study is that the computational burden will increase substantially as the number of BESS in the power system increases. Although the proposed L-BD-SCUC method can solve the cases with multi-BESSs, the low efficiency indicates that further research beyond this work and our prior work is still needed. To summarize, with the proposed L-BD-SCUC, an alternative method is available to efficiently solve the NNBD embedded SCUC problem.